\documentclass{ifacconf}

\usepackage{graphicx}      
\usepackage{natbib}        

\usepackage{graphicx} 
\usepackage{amsmath} 
\usepackage{amssymb}  

\usepackage{bm}
\usepackage{bbm}
\usepackage{mathtools}
\mathtoolsset{showonlyrefs}
\usepackage{upgreek}
\usepackage{accents}
\usepackage{bbold}

\usepackage{pifont}

\newcommand{\mf}{\mathbf}

\newcommand{\cov}{\normalfont\textsf{\footnotesize cov}}

\newcommand{\ubar}[1]{\underaccent{\bar}{#1}}
\usepackage{verbatim}
\usepackage{multirow}
\usepackage{xcolor}

\newtheorem{proposition}{Proposition}

\newtheorem{assumption}{Assumption}
\newtheorem{remark}{Remark}

\usepackage{paralist}
\usepackage{booktabs}

\begin{document}
\begin{frontmatter}

\title{Event-Triggered Consensus for Continuous-Time Distributed Estimation \thanksref{footnoteinfo}}

\thanks[footnoteinfo]{This work was supported via projects PID2021-124137OB-I00 and TED2021-130224B-I00 funded by MCIN/AEI/10.13039/501100011033, by ERDF A way of making Europe and by the European Union NextGenerationEU/PRTR, by the Gobierno de Aragón under Project DGA T45-20R, by the Universidad de Zaragoza and Banco Santander, by the Consejo Nacional de Ciencia y Tecnología (CONACYT-México) grant 739841, and by Spanish grant FPU20/03134.
{\color{red} \copyright 2023 IFAC. This work has been accepted to IFAC for publication under a Creative Commons Licence CC-BY-NC-ND. Accepted for presentation at the 22nd IFAC World Congress 2023.}}

\author[First]{Irene Perez-Salesa} 
\author[First]{Rodrigo Aldana-Lopez} 
\author[First]{Carlos Sagues}

\address[First]{Departamento de Informática e Ingeniería de Sistemas (DIIS) and Instituto de Investigación en Ingeniería de Aragón (I3A), Universidad de Zaragoza, María de Luna 1, 50018 Zaragoza, Spain}

\begin{abstract}                
Distributed sensor networks have gained interest thanks to the developments in processing power and communications. Event-triggering mechanisms can be useful in reducing communication between the nodes of the network, while still ensuring an adequate behaviour of the system. However, very little attention has been given to continuous-time systems in this context. In this work, we propose a strategy for distributed state estimation in sensor networks, based on average dynamic consensus of the continuous measurements. While communication between nodes is discrete and heavily reduced due to the event-triggering mechanism, our method ensures that the nodes are still able to produce a continuous estimate of the global average measurement and the state of the plant, within some tuneable error bounds.  
\end{abstract}


\end{frontmatter}

\section{Introduction}

Distributed sensor networks have gained interest thanks to the developments in processing power and communications. In distributed state estimation problems, several sensors collectively observe a dynamical system. Each of them has access to a local measurement as well as  information from neighboring sensors, which can be used to reconstruct the full state of the plant. The collaboration of several sensing agents has some benefits. First, as it is pointed out by \cite{Ren2018}, the redundancy of having several sensors measuring the same variables results in less risk of single-point failure, as well as a decrease in uncertainty of the resulting estimate. Additionally, since the nodes share information with their neighbors, it is no longer needed that each sensing node can estimate the full state of the plant using exclusively its own local measurement. Thus, several works tackle the problem of distributed state estimation, proposing distributed implementations of well-known filters \citep{Olfati-Saber2005, George2013, Ren2018}.

Distributed estimation comes at the cost of having communication between the elements of the sensor network. This communication can be constrained by aspects such as bandwidth, a shared transmission network that needs to be available to several elements, or power limitations of the nodes. For this reason, event-triggered strategies are of interest for distributed applications, such as in networked systems \citep{Miskowicz2014} or wireless sensor networks \citep{Jia2021}. 
In this context, event-triggering mechanisms need to be chosen to reduce transmissions of information, while ensuring that the quality of the resulting estimates is not heavily degraded. 

Several approaches have been taken to achieve distributed state estimation with event-triggered communication. Due to the availablity of a local estimate in each sensing node, some works design their triggering condition based on their local estimate of the state \citep{Battistelli2018, Yu2020}. Another option is to monitor the behaviour of the measured signal, applying an absolute threshold to the difference between the current and last transmitted value \citep{Ge2019, Ding2020, Li2021a, Zhu2022}, or to base the trigger on the innovation of the measurements, i.e. the difference between the real and predicted measurement \citep{Liu2015, Yang2017, Qian2021}. To fuse the information, adequate state estimators are designed according to the triggering mechanism, or algorithms to reach consensus on the state estimates are used \citep{Meng2014, He2020, Rezaei2022}. While a variety of works exist on this topic, all of the aforementioned ones focus on discrete-time systems. Very little attention has been given to continuous-time systems in this context. Moreover, works such as \cite{Ding2015,Zhang2017}, which do feature continuous-time systems, do not consider stochastic noise.

Motivated by this discussion, we contribute a novel setup to achieve distributed state estimation in continuous-time systems, based on dynamic average consensus of the measurements with event-triggered communication between nodes. Contrary to other works, in which consensus is done on the state estimates, each node has a triggering condition based on its local estimate of the global average measurement. Thus, events are triggered according to the evolution of the measured variables. While nodes communicate their local consensus estimates to their neighbors in a discrete manner due to the event-triggering mechanism, our method ensures that each node is still able to estimate the continuous global measurement within an error threshold that can be tuned by the user. To the best of our knowledge, this is the first time event-triggered communication has been used for distributed estimation of continuous time plants and measurements with stochastic noise. We show the effectiveness of our proposal in reducing transmissions compared to an ideal case with continuous communication between nodes, while still achieving a comparable estimation performance.

\subsection{Notation}
We denote the $n \times n$ identity matrix as $\mf{I}_n$. Moreover, let $\mathbb{1}=[1,\dots,1]^\top$ for an appropriate size. The Euclidean norm is represented by $\|\bullet\|$. Let $\otimes$ denote the Kronecker product.

\section{Problem Statement}
\label{sec:problem}
Consider an unknown input dynamical system of the form
\begin{equation}
\label{eq:system_sde}
    \dot{\mf{x}}(t) = \mf{A}\mf{x}(t) + \mf{B}\mf{w}(t), \ \ t \geq 0
\end{equation}
where $\mf{A}\in\mathbb{R}^{n \times n},\; \mf{B}\in\mathbb{R}^{{n}\times {n_\mf{w}}}$ and $\mf{w}(t)\in\mathbb{R}^{n_{\mf{w}}}$ is the unknown input, which can also contain disturbances or other non-modeled dynamics. In order to apply optimal filtering techniques such as a Kalman filter to \eqref{eq:system_sde}, $\mf{w}(t)$ is usually modelled as an ${n_\mf{w}}$-dimensional Wiener process with $\cov\{\mf{w}(s),\mf{w}(r)\} = \mf{W}\min(s,r)$ \cite[Page 63]{astrom}. In this case, \eqref{eq:system_sde} is better understood as a Stochastic Differential Equation (SDE), with $\mf{x}(t)$ following a Gaussian distribution where the mean ${\mf{x}}_0$ and covariance matrix $\mf{P}_0$ for the initial condition $\mf{x}(0)$ are assumed to be known.

The state of the plant is monitored by a sensor network composed of $N$ sensors. The communication network topology is modeled by a connected undirected graph $\mathcal{G}$, with node set $\mathcal{V} = \{1,\dots,N\}$ for the sensors and an edge set $\mathcal{E}\subseteq \mathcal{V}\times\mathcal{V}$ for the communication links between neighboring nodes. The adjacency matrix  $\mathbf{A}_{\mathcal{G}}=[a_{ij}]\in\{0,1\}^{N\times N}$ has elements $a_{ij} = 1$ if $(i,j) \in \mathcal{E}$ and $a_{ij} = 0$ otherwise. The set of neighbors of node $i$ is denoted by $\mathcal{N}_i = \{j \in \mathcal{V} : (i,j) \in \mathcal{E}\}$.

Each node $i$ has access to its own measurements $\mf{y}_i(t) = \mf{C}_i \mf{x}(t) + \mf{v}_i(t), \forall t \geq 0$, where $\mf{C}_i \in \mathbb{R}^{n_\mf{y} \times n}$ is a constant matrix. Moreover, $\mf{v}_i(t)$ is modeled to follow a Gaussian distribution with zero mean and covariance matrix $\mf{R}_i$. Assume that the pair $(\mf{A}, [\mf{C}_1^\top,\dots,\mf{C}_N^\top]^\top)$ is observable.

Communication between sensors is triggered at each node according to a condition based on the node's local information. The goal for each node of the system is to compute an estimate of the full state of the plant, by fusing its own local information and the transmitted information from neighboring nodes. 

\section{Distributed Estimation under Event-Triggered Communication}

\begin{figure}
    \centering
    \includegraphics[width=\columnwidth]{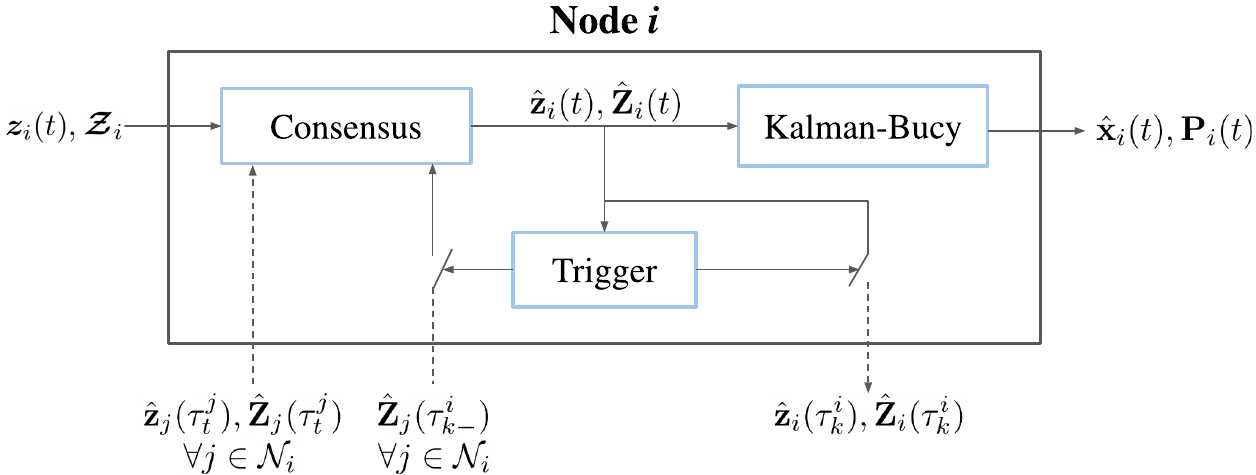}
    \caption{Summary of the proposal. Using a consensus algorithm, node $i$ computes estimates $\hat{\mf{z}}_i(t)$ and $\hat{\mf{Z}}_i(t)$ of the average information vector $\bar{\mf{z}}(t)$ and matrix $\bar{\mf{Z}}$, respectively. The node uses its local measurement information, given by $\bm{z}_i(t)$ and $\bm{\mathcal{Z}}_i$, as well as estimates from its neighbors $j \in \mathcal{N}_i$. The consensus results are used in a Kalman-Bucy filter to estimate the state of the plant. An event trigger decides when to exchange information with neighbors. Dashed lines indicate event-triggered communication between nodes.}
    \label{fig:summary}
\end{figure}

We propose a solution to distributed state estimation in sensor networks with communication constraints via dynamic consensus of the measurements. Each node uses its local information and that of its neighbors to compute an average measurement through a consensus algorithm. In order to reduce communication between nodes, we equip each sensor with an event-triggering mechanism to decide when to broadcast its local information to the neighbors. Even though communication between nodes is performed at discrete event instants, the sensors have access to a local continuous measurement and the consensus algorithm is also run in a continuous fashion. Thus, each node computes a \emph{continuous} average measurement, which is fed to a Kalman-Bucy filter to produce an estimate of the state in each node. 
Our proposal is summarized in Figure \ref{fig:summary}.

\subsection{Event-Triggered Consensus of Measurements}

First, let the informational form of the measurement $\bm{z}_i(t) := \mf{C}_i^\top\mf{R}_i^{-1}\mf{y}_i(t)$ with information matrix $\bm{\mathcal{Z}}_i :=\mf{C}_i^\top\mf{R}_i^{-1} \mf{C}_i$ and define the following consensus quantities:
\begin{equation}\label{eq:consensus_quantities}
    \begin{aligned}
    \bar{\mf{z}}(t) &:= \frac{1}{N} \sum_{i=1}^N \bm{z}_i(t),\ \ \bar{\mf{Z}}:= \frac{1}{N} \sum_{i=1}^N \bm{\mathcal{Z}}_i
    \end{aligned}
\end{equation}

Each node has access to its local values of $\bm{z}_i(t)$ and $\bm{\mathcal{Z}}_i$, and computes the estimates $\hat{\mf{z}}_i(t)$ and $\hat{\mf{Z}}_i(t)$ for the consensus quantities $\bar{\mf{z}}(t)$ and $\bar{\mf{Z}}$ using only local communication.

We consider that each node $i$ communicates with its neighbors only at some event instants $t \in \{\tau_k^i\}_{k=0}^\infty$. These events are constructed by the following absolute threshold triggering condition, applied to the local estimate $\hat{\mf{z}}_i(t)$:
\begin{equation}\label{eq:trigger}
    \tau_{k+1}^i = \inf\{ t-\tau_k^i>\ubar{\tau}\ \ | \ \ \|\hat{\mf{z}}_i(t) - \hat{\mf{z}}_i(\tau_k^i)\| \geq \delta_i \} \\
\end{equation}
where $\ubar{\tau}>0$ is included as time regularization in order to guarantee a minimum inter-event time and $\delta_i > 0$ is a design parameter. 

\begin{remark}
Note that $\hat{\mf{z}}_i(t)$ evolves according to the measurements, with the goal of reaching a consensus on the average measurement of the plant. Thus, setting the triggering condition on this variable means that communication can be greatly reduced when the measured signals do not suffer significant changes.
\end{remark}

When an event is triggered at node $i$, the node broadcasts its value of $\hat{\mf{z}}_i(t)$ to its neighbors. Thus, node $i$ has knowledge of its own measurement $\bm{z}_i(t)$ and local estimate $\hat{\mf{z}}_i(t)$, as well as the last transmitted value from its neighbors, $\hat{\mf{z}}_j(\tau^j_t) \forall j \in \mathcal{N}_i$, where $\tau^{j}_t=\max\{\tau_k^j\leq t\}$ is the last triggering instant at node $j$ prior to time $t$. The proposed consensus algorithm to compute $\hat{\mf{z}}_i(t)$ at each node can be expressed as:
\begin{equation}\label{eq:ev_consensus}
\begin{aligned}
    \dot{\mf{p}}_i(t) &= - \kappa_1 \mf{p}_i(t) + \kappa_2 \sum_{j \in \mathcal{N}_i}  \left( \hat{\mf{z}}_i(t) - \hat{\mf{z}}_j(\tau^{j}_t) \right) \\
    \hat{\mf{z}}_i(t) &= \bm{z}_i(t)-\mf{p}_i(t)
\end{aligned}
    \end{equation}
where $\mf{p}_i(t)$ is an auxiliary local variable and $\kappa_1, \kappa_2$ are design parameters.

In order to compute $\bar{\mf{Z}}$ from each node, we propose a scheme of discrete updates at event instants. First, let $\hat{\mf{Z}}_i(t)$ be the estimate of $\bar{\mf{Z}}$ at node  $i$ for any $t\geq 0$ which complies $\hat{\mf{Z}}_i(0) = \bm{\mathcal{Z}}_i$. The estimate $\hat{\mf{Z}}_i(t)$ is piece-wise constant and changes its value only when node $i$ or its neighbors $j\in\mathcal{N}_i$ trigger an event, according to the following rules. When an event $t=\tau_{k}^i$ is triggered in node $i$ due to \eqref{eq:trigger}, the node asks its neighbors $j \in \mathcal{N}_i$ for their last updated estimates $\hat{\mf{Z}}_j(\tau_{k-}^i) := \lim_{t\to(\tau_k^i)^-}\hat{\mf{Z}}_j(t)$. Then, node $i$ updates its value as 
\begin{equation}\label{eq:Z_update}
    \hat{\mf{Z}}_i(\tau_{k}^i) = \frac{\hat{\mf{Z}}_i(\tau_{k-}^i) + \sum_{j\in \mathcal{N}_i} \hat{\mf{Z}}_j(\tau_{k-}^i)} {J_i + 1}
\end{equation}
where $J_i:=\sum_{j=1}^N a_{ij}$ denotes the number of neighbors of node $i$. Node $i$ broadcasts $\hat{\mf{Z}}_i(\tau_k^i)$ at $t=\tau_k^i$ to its neighbors, which update their estimate as:
\begin{equation}\label{eq:Z_neighbors}
    \begin{aligned}
        \hat{\mf{Z}}_j(\tau_{k}^i) &= \hat{\mf{Z}}_i(\tau_{k}^i) \,\,\, &\forall j \in \mathcal{N}_i
    \end{aligned}
\end{equation}
Moreover, $\hat{\mf{Z}}_i(t)$ remains constant between event instants.

\subsection{Distributed State Estimation}

The resulting continuous consensus quantities computed at each node can be incorporated to a Kalman-Bucy filter in order to obtain an estimate of the state of the plant. The distributed implementation of the filter, taking into account the consensus quantities in \eqref{eq:consensus_quantities}, can be expressed as \citep{Ren2018}:
\begin{equation}\label{eq:kalman-bucy}
\begin{aligned}
    \dot{\hat{\mf{x}}}_i(t) =& \mf{A}\hat{\mf{x}}_i(t) + N\mf{P}_i(t)\hat{\mf{z}}_i(t) - N\mf{P}_i(t)\hat{\mf{Z}}_i(t)\hat{\mf{x}}_i(t) \\
    \dot{\mf{P}}_i(t)=& \mf{AP}_i(t) + \mf{P}_i(t)\mf{A}^\top + \mf{BWB}^\top - N \mf{P}_i(t)\hat{\mf{Z}}_i(t)\mf{P}_i(t) \\
\end{aligned}
\end{equation}
where the network size $N$ is assumed to be known
either by construction, or obtained through well-known distributed leaderless methods as in \cite{network_size_estimation}. 
\begin{remark}
Note that, since we are using an event-triggered implementation of the consensus algorithm, an additional error is introduced in the consensus phase with respect to an ideal case with continuous communication between nodes. Feeding the consensus signal directly to the Kalman-Bucy filter is an approximation, which does not take the additional event-triggered error into account. However, as we show in the following section, the error due to events can be made arbitrarily small by tuning the triggering thresholds $\delta_i$, the constants in \eqref{eq:ev_consensus} or by improving the connectivity of the network. 
\end{remark}

\section{Main Results}
In order to show that the consensus filter in \eqref{eq:ev_consensus} works, we use the following assumption.
\begin{assumption}\label{as:bound_u}
Given $\kappa_1$ in \eqref{eq:ev_consensus}, there exists a bound $L$ such that $\|\dot{\bar{\mf{z}}}(t)-\dot{\bm{z}}_i(t) - \kappa_1({\bar{\mf{z}}}(t)-{\bm{z}}_i(t))\| \leq L$, for all $i\in\mathcal{V}, \forall t\geq 0$.
\end{assumption}
Assumption \ref{as:bound_u} is reasonable in practice and similar assumptions have been made previously in the literature \citep{Olfati-Saber2005,Ren2018}.
\begin{proposition}\label{prop:error-z}
Let $\mathcal{G}$ be a connected graph with algebraic connectivity $\lambda_2(\mathcal{G})$, Assumption \ref{as:bound_u} hold, the event-triggering rule \eqref{eq:trigger} and the consensus algorithm in \eqref{eq:ev_consensus} with parameters $\{\delta_i\}_{i=1}^N,\kappa_1,\kappa_2$. Hence, there exist constants $T, K>0$ such that
$$
\|\bar{\mf{z}}(t)-\hat{\mf{z}}_i(t)\|\leq K , \forall t\geq T,
$$
where $K$ can be made arbitrarily small by decreasing $\max\{\delta_i\}_{i=1}^N$ and increasing $\kappa_1$, $\kappa_2, \kappa_1/\kappa_2$.
\end{proposition}
\begin{pf}
Define $\mf{u}_{j}(t) = \hat{\mf{z}}_j(t) - \hat{\mf{z}}_j(\tau_t^{j})$ with arbitrary $j\in\mathcal{N}_i$. Then, the consensus algorithm in \eqref{eq:ev_consensus} can be expressed as
\begin{equation}
    \begin{aligned}\label{eq:ev_consensus_w}
        \dot{\mf{p}}_i(t) = -\kappa_1 \mf{p}_i(t) + \kappa_2 \sum_{j \in \mathcal{N}_i}  (\hat{\mf{z}}_i(t) - \hat{\mf{z}}_j(t)) + \kappa_2\sum_{j\in \mathcal{N}_i} \mf{u}_{j}(t)
    \end{aligned}
\end{equation}
Now, let: 
\begin{equation}
\begin{aligned}
    &\mf{p}(t) = \begin{bmatrix}\mf{p}_1(t) \\ \vdots \\ \mf{p}_N(t) \end{bmatrix}, \;
    \hat{\mf{z}}(t) = \begin{bmatrix}\hat{\mf{z}}_1(t) \\ \vdots \\ \hat{\mf{z}}_N(t) \end{bmatrix} \\
    &\bm{z}(t) = \begin{bmatrix}\bm{z}_1(t) \\ \vdots \\ \bm{z}_N(t) \end{bmatrix}, \;
    \mf{u}(t) = \begin{bmatrix}\mf{u}_1(t) \\ \vdots \\ \mf{u}_N(t) \end{bmatrix}
\end{aligned}
\end{equation} 
used in order to write \eqref{eq:ev_consensus_w} in matrix form as
\begin{equation}
    \dot{\mf{p}}(t) = -\kappa_1 \mf{p}(t) + \kappa_2( \mathbf{Q}_{\mathcal{G}}\otimes\mf{I}_{n})\hat{\mf{z}}(t) + \kappa_2 (\mathbf{A}_{\mathcal{G}}\otimes \mf{I}_n)\mf{u}(t)
\end{equation}
where $\mathbf{Q}_{\mathcal{G}}$ and $\mathbf{A}_{\mathcal{G}}$ denote the Laplacian and adjacency matrices of the graph $\mathcal{G}$ \citep{Godsil}. Let $\bar{\mf{s}}(t) = (\mathbb{1}^\top/N \otimes \mf{I}_n)\hat{\mf{z}}(t)$ and $\tilde{\mf{s}}(t) = (\mf{H} \otimes \mf{I}_n)\hat{\mf{z}}(t)$ be the consensus component and consensus error respectively with $\mf{H}=\mf{I}_N - (1/N)\mathbb{1}\mathbb{1}^\top$. Note that $\hat{\mf{z}}(t) = (\mathbb{1}\otimes \mf{I}_n)\bar{\mf{s}}(t) + \tilde{\mf{s}}(t)$.
Moreover, the dynamics of $\tilde{\mf{s}}(t)$ comply:
\begin{equation}
\begin{aligned}
&\dot{\tilde{\mf{s}}} = (\mf{H} \otimes \mf{I}_n)\dot{\hat{\mf{z}}} = (\mf{H} \otimes \mf{I}_n)(\dot{\bm{z}} - \dot{{\mf{p}}}) \\
&= (\mf{H} \otimes \mf{I}_n)(\dot{\bm{z}} + \kappa_1 (\bm{z} - \hat{\mf{z}}) - \kappa_2 (\mf{Q}_\mathcal{G}\otimes \mf{I}_n)\hat{\mf{z}} \\
&- \kappa_2 (\mf{A}_\mathcal{G}\otimes \mf{I}_n) \mf{u}) \\
&= (\mf{H} \otimes \mf{I}_n)(\dot{\bm{z}} + \kappa_1 \bm{z}) - \kappa_1(\mf{H} \otimes \mf{I}_n)\hat{\mf{z}} \\&- \kappa_2(\mf{H} \otimes \mf{I}_n)(\mf{Q}_\mathcal{G} \otimes \mf{I}_n)\hat{\mf{z}} - \kappa_2 (\mf{H} \otimes \mf{I}_n)(\mf{A}_\mathcal{G} \otimes \mf{I}_n) \mf{u} \\
&=\tilde{\bm{z}} - \kappa_1\tilde{\mf{s}} - \kappa_2 (\mf{Q}_\mathcal{G} \otimes \mf{I}_n)\tilde{\mf{s}} - \kappa_2 (\mf{H}\mf{A}_\mathcal{G} \otimes \mf{I}_n)\mf{u} \\
\end{aligned}
\end{equation}
omitting time dependency for brevity and defining $\tilde{\bm{z}}(t) = (\mf{H} \otimes \mf{I}_n)(\dot{\bm{z}}(t) + \kappa_1 \bm{z}(t))$ which complies $\|\tilde{\bm{z}}(t)\|\leq L'$ for some $L'>0$ due to Assumption \ref{as:bound_u}. Define the Lyapunov function candidate $\mf{V}(\tilde{\mf{s}}(t)) = \tilde{\mf{s}}(t)^\top \tilde{\mf{s}}(t)$. Then, we have
\begin{equation}
\begin{aligned}
    &\dot{\mf{V}}(\tilde{\mf{s}}) = 2 \tilde{\mf{s}}^\top \dot{\tilde{\mf{s}}} \\
    &= 2 \tilde{\mf{s}}^\top \tilde{\bm{z}} - 2\kappa_1 \tilde{\mf{s}}^\top \tilde{\mf{s}} - 2\kappa_2 \tilde{\mf{s}}^\top (\mf{Q}_\mathcal{G} \otimes \mf{I}_n)\tilde{\mf{s}}\\& - 2 \kappa_2 \tilde{\mf{s}}^\top (\mf{H}\mf{A}_\mathcal{G} \otimes \mf{I}_n)\mf{u}\\
    &\leq 2 \|\tilde{\mf{s}}\| \|\tilde{\bm{z}}\| - 2\kappa_1 \|\tilde{\mf{s}}\|^2 - 2 \kappa_2 \lambda_2(\mathcal{G}) \|\tilde{\mf{s}}\|^2 \\&+ 2 \kappa_2 \|\tilde{\mf{s}}\| \|(\mf{HA}_\mathcal{G} \otimes \mf{I}_n)\mf{u}\| \\
    &\leq 2 \|\tilde{\mf{s}}\| ( \|\tilde{\bm{z}}\| - \kappa_2 \lambda_2(\mathcal{G}) \|\tilde{\mf{s}}\| + \kappa_2 \|(\mf{HA}_\mathcal{G} \otimes \mf{I}_n)\mf{u}\|)
    \end{aligned}
\end{equation}
with $\lambda_2(\mathcal{G})$ being the connectivity of the network, i.e. the minimum nonzero eigenvalue of $\mf{Q}_\mathcal{G}$. Note that the eigenvalues of $\mf{Q}_\mathcal{G} \otimes \mf{I}_n$ coincide with those of $\mf{Q}_\mathcal{G}$ with additional multiplicity by $n$. This yields that $\dot{\mf{V}}(\tilde{\mf{s}}(t))<0$ when 
\begin{equation}
    \|\tilde{\mf{s}}\| > \frac{\|\tilde{\bm{z}}\| + \kappa_2\|(\mf{HA}_\mathcal{G} \otimes \mf{I}_n) \mf{u}\| }{\kappa_2 \lambda_2(\mathcal{G})} 
\end{equation}
Hence, for any initial condition there exists $T>0$ such that $\tilde{\mf{s}}(t)$ converges to the region in which 
 $\|\tilde{\mf{s}}(t)\|\leq \tilde{K}, \forall t\geq T$ where
$$
\tilde{K} = \frac{L' + \kappa_2\sigma_{\max}(\mf{HA}_\mathcal{G} \otimes \mf{I}_n)\sqrt{N}\max\{\delta_i\}_{i=1}^N  }{\kappa_2 \lambda_2(\mathcal{G})} 
$$
with $\sigma_{\max}(\bullet)$ denoting the maximum singular value and where we used $\|(\mf{HA}_\mathcal{G} \otimes \mf{I}_n) \mf{u}(t)\|\leq\sigma_{\max}(\mf{HA}_\mathcal{G} \otimes \mf{I}_n)\|\mf{u}(t)\|$ with $\|\mf{u}_i(t)\| \leq \delta_i$ and $\| \tilde{\bm{z}} \|\leq L'$.

Finally, it remains to check that $\bar{\mf{s}}(t)$ converges to a neighborhood of $\bar{\mf{z}}(t)$.
From the dynamics of $\bar{\mf{s}}(t)$, we have the following:
\begin{equation}
    \begin{aligned}
    &\dot{\bar{\mf{s}}} = \left( {\mathbb{1}^\top}/N \otimes \mf{I}_n \right)\dot{\hat{\mf{z}}} \\
    &= \left({\mathbb{1}^\top}/N\otimes \mf{I}_n \right)(\dot{\bm{z}} + \kappa_1 (\bm{z} - \hat{\mf{z}}) \\&- \kappa_2 (\mf{Q}_\mathcal{G}\otimes \mf{I}_n)\hat{\mf{z}} - \kappa_2 (\mf{A}_\mathcal{G}\otimes \mf{I}_n) \mf{u})\\
    &= \dot{\bar{\mf{z}}} + \kappa_1 (\bar{\mf{z}} - \bar{\mf{s}}) - \kappa_2 \left({\mathbb{1}^\top} \mf{A}_\mathcal{G}/N \otimes \mf{I}_n\right) \mf{u}
    \end{aligned}
\end{equation}

Defining the error $\mf{e}(t) = \bar{\mf{s}}(t) - \bar{\mf{z}}(t)$ and disturbance $\bar{\mf{u}}(t) = - \kappa_2 \left(\mathbb{1}^\top\mf{A}_\mathcal{G}/N \otimes \mf{I}_n\right) \mf{u}(t)$, it follows that $\dot{\mf{e}}(t) = -\kappa_1 \mf{e}(t) + \bar{\mf{u}}(t)$, which has the following explicit solution:
\begin{equation}
    \mf{e}(t) = \mf{e}(0)e^{-\kappa_1 t} + e^{-\kappa_1 t} \int_{0}^{t}e^{\kappa_1 \tau} \bar{\mf{u}}(\tau) \d \tau
\end{equation}
Hence, for any $t\geq T$ we have that
$$
\begin{aligned}
&\|\mf{e}(t)\|\leq e^{-\kappa_1 T}\|\mf{e}(0)\| + e^{-\kappa_1 T}\int_0^T e^{\kappa_1 \tau}\d \tau \left(\sup_{\tau\geq 0} \|\bar{\mf{u}}(\tau)\| \right) \\
&\leq e^{-\kappa_1 T}\|\mf{e}(0)\| + \frac{\kappa_2}{\kappa_1}\sigma_{\max}\left(\mathbb{1}^\top\mf{A}_\mathcal{G}/N \otimes \mf{I}_n\right)\sqrt{N}\max\{\delta\}_{i=1}^N \\
&=: \bar{K}
\end{aligned}
$$
Hence, for $t\geq T$ the consensus error $\tilde{\mf{s}}(t)$ is bounded by $\tilde{K}$ which decreases with respect to $\kappa_1,\kappa_2,\lambda_2(\mathcal{G}),\max\{\delta\}_{i=1}^N$ as pointed out in the theorem statement. Similarly, the consensus component error $\mf{e}(t)=\bar{\mf{s}}(t) - \bar{\mf{z}}(t)$ is bounded by $\bar{K}$ which decreases in a similar fashion. Hence, $\|\bar{\mf{z}}(t)-\hat{\mf{z}}_i(t)\|, \forall t\geq T$ is bounded by a constant $K$ which takes into account the effect of $\tilde{K}, \bar{K}$. \qed
\end{pf}

\begin{proposition}\label{prop:error_S}
Let $\mathcal{G}$ be a connected graph and consider the event-triggering mechanism in \eqref{eq:trigger} along with the consensus algorithm in \eqref{eq:Z_update} and \eqref{eq:Z_neighbors}. Then, $\hat{\mf{Z}}_i(t)$ asymptotically converges to $\bar{\mf{Z}}=\sum_{i=1}^N \bm{\mathcal{Z}}_i/N$ as events occur. 
\end{proposition}

\begin{pf}
Consider the global sequence of events in all nodes as the overlapping sequence $\{\tau_k\}_{k=1}^\infty = \bigcup_{i=1}^{N} \{\tau_k^i\}_{k=1}^\infty$. Without loss of generality, we assume that any $\tau_k$ corresponds to the event from a single node.
Note that \eqref{eq:Z_update} shows that $\hat{\mf{Z}}_i(\tau_{k+1})$ is computed via a convex combination of $\hat{\mf{Z}}_i(\tau_k)$ and $\hat{\mf{Z}}_j(\tau_k)\, \forall j \in \mathcal{N}_i$ with equal weights $\lambda = 1/(J_i +1)$, complying $\lambda + \sum_{j\in\mathcal{N}_i} \lambda = 1$. 

Choose an arbitrary component ${s}_i(\tau_k)$ of the matrix $\hat{\mf{Z}}_i(\tau_k)$ and let $\mf{s}(\tau_k) = [{s}_1(\tau_k), \dots, {s}_N(\tau_k)]^\top$. Then, define the Lyapunov function candidate $\mf{V}(\mf{s}) = \textsf{max}(\mf{s}) - \textsf{min}(\mf{s})$ and note that $\mf{V}(\mf{s}) = 0$ only if $\textsf{max}(\mf{s}) = \textsf{min}(\mf{s})$, i.e. when ${s}_i = {s}_j,  \forall i,j \in \mathcal{V}$. 
To show convergence of the elements of $\mf{s}(\tau_k)$, the Lyapunov function must be non-increasing. We have
\begin{equation}\label{eq:lyap_decrease}
\begin{aligned}
    &\mf{V}(\mf{s}(\tau_{k+1})) - \mf{V}(\mf{s}(\tau_k)) = \textsf{max}({\mf{s}}(\tau_{k+1})) - \textsf{max}({\mf{s}}(\tau_{k})) \\ &- (\textsf{min}({\mf{s}}(\tau_{k+1})) - \textsf{min}(\mf{s}(\tau_{k})))
\end{aligned}
\end{equation}
Note that, since the elements of $\mf{s}(\tau_{k+1})$ are computed through a convex combination of elements of $\mf{s}(\tau_{k})$, they are contained in their convex hull. Hence, it follows that
\begin{equation}\label{eq:s_bounds}
\begin{aligned}
    \textsf{min}({\mf{s}}(\tau_{k})) \leq \textsf{min}({\mf{s}}(\tau_{k+1})) \leq 
    \textsf{max}({\mf{s}}(\tau_{k+1})) \leq \textsf{max}({\mf{s}}(\tau_{k}))
\end{aligned}
\end{equation}
which shows that $\mf{V}(\mf{s}(\tau_{k+1})) - \mf{V}(\mf{s}(\tau_k))\leq 0$ holds and the component ${s}_i(\tau_k)$ of the matrix $\hat{\mf{Z}}_i(\tau_k)$ for all nodes asymptotically converge to the same value i.e., $\lim_{k\to\infty}s_i(\tau_k)=\alpha, \forall i\in\mathcal{V}$  for some $\alpha \geq 0$.  Moreover, note that:
\begin{equation}
\begin{aligned}
    &\sum_{\ell=1}^N s_\ell(\tau_{k+1}) = \sum_{\ell\neq i,j\in\mathcal{N}_i}s_\ell(\tau_{k+1}) + s_i(\tau_{k+1}) + \sum_{j\in\mathcal{N}_i}s_j(\tau_{k+1})\\ 
    &=\sum_{\ell\neq i,j\in\mathcal{N}_i}s_\ell(\tau_{k}) + (J_i+1)s_i(\tau_{k+1}) \\
    &=\sum_{\ell\neq i,j\in\mathcal{N}_i}s_\ell(\tau_{k}) + \left(s_i(\tau_{k}) + \sum_{\ell\in\mathcal{N}_i}s_\ell(\tau_{k})\right)=\sum_{\ell=1}^N s_\ell(\tau_k) 
\end{aligned}
\end{equation}
where the update $$
s_i(\tau_{k+1})=s_j(\tau_{k+1})=\frac{s_i(\tau_k)+\sum_{j\in\mathcal{N}_i}s_j(\tau_k)}{J_i+1}$$ 
from \eqref{eq:Z_update} and \eqref{eq:Z_neighbors} was used.
Hence, $\sum_{i=1}^N s_i(\tau_{k}) = \sum_{i=1}^N s_i(0)$ remains invariant $\forall k\geq 0$.
Moreover, the consensus result implies $\lim_{k\to\infty} \sum_{i=1}^N s_i(\tau_k) = \alpha N$. Therefore, it must be the case that $\alpha = \sum_{i=1}^N s_i(0)/N$. Thus, all nodes converge to the average of the initial conditions. This reasoning can be extended to all elements of the matrix $\hat{\mf{Z}}_i(\tau_k)$, showing that all nodes reach the global average value $\bar{\mf{Z}} = \sum_{i=1}^N \hat{\mf{Z}}_i(0)/N=\sum_{i=1}^N \bm{\mathcal{Z}}_i/N$. \qed
\end{pf}

\section{Simulation Experiments}

\begin{figure}
    \centering
    \includegraphics[width=0.4\columnwidth]{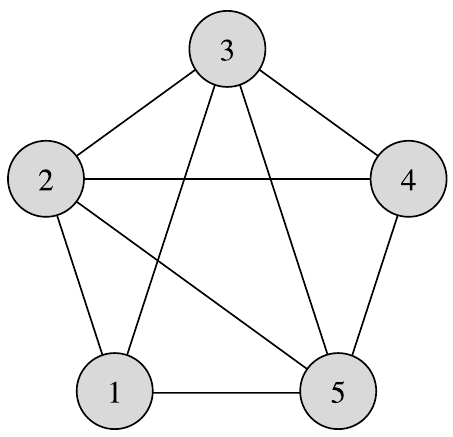}
    \caption{Graph $\mathcal{G}$ describing the sensor network.}
    \label{fig:graph}
\end{figure}

Consider a 2-D object tracking problem. Let the state vector $\mf{x}(t) = [x(t), y(t), v_x(t), v_y(t)]$ of the object where $(x(t), y(t))$ represent Cartesian coordinates and the corresponding velocity components for both axis are represented by $(v_x(t), v_y(t))$. The object moves in the following trajectory, which is unknown to the observers:
\begin{equation}
\label{eq:trajectory}
    \mf{x}(t) = \begin{bmatrix}
    \textsf{sin}(0.5t)\\
    3.5\textsf{sin}(0.8t)\\
    0.5\textsf{cos}(0.5t)\\
    2.8\textsf{cos}(0.8t)\\
    \end{bmatrix}
\end{equation}
For this experiment to be realistic, the trajectory in \eqref{eq:trajectory} is not a stochastic process. However, in absence of knowledge of the unknown input, nodes model \eqref{eq:trajectory} conservatively by the SDE in \eqref{eq:system_sde}, with
\begin{equation}
\mf{A} =   \begin{bmatrix}
		0 & 0 & 1 & 0\\	
		0 & 0 & 0 & 1\\
		0 & 0 & 0 & 0\\
		0 & 0 & 0 & 0\\
		\end{bmatrix}, \;
\mf{B} =   \begin{bmatrix}
		0 & 0\\	
		0 & 0\\
		1 & 0\\
		0 & 1\\
		\end{bmatrix}
\end{equation}
and $\mf{W} = \textsf{diag}(1, 1)$. At $t=0$, the state is modeled by the sensors with a Gaussian distribution with mean $\mf{x}_0 = [0,0,0.5,2.8]^\top$ and covariance $\mf{P}_0 = \mf{I}_{n}$.
The system is observed by a sensor network consisting of $N=5$ nodes, as shown in Figure \ref{fig:graph}. Each of them can access a local measurement $\mf{y}_i(t) = \mf{C}_i\mf{x}(t) + \mf{v}_i(t)$, with
\begin{equation}
\begin{aligned}
&\mf{C}_1 = \mf{C}_5 =  \begin{bmatrix}
		1 & 0 & 0 & 0\\	
		\end{bmatrix}, \,
\mf{C}_3 = \begin{bmatrix}
        0 & 1 & 0 & 0 \\
        \end{bmatrix} \\
&\mf{C}_2 = \mf{C}_4 =  \begin{bmatrix}
		1 & 0 & 0 & 0\\
		0 & 1 & 0 & 0
		\end{bmatrix}
\end{aligned}
\end{equation}
and noise covariances $\mf{R}_1 = 0.02, \, \mf{R}_3 = 0.01, \, \mf{R}_5 = 0.015, \mf{R}_2 = \mf{R}_4 = \textsf{diag}(0.01, 0.01)$. For the event-triggered simulations, we have set the same triggering threshold for all nodes. The constants in the consensus algorithm \eqref{eq:ev_consensus} have been set to $\kappa_1 = 0.5, \, \kappa_2 = 20$. The simulation time has been set to $T_f = 10$ with a step of $h=1\cdot 10^{-4}$.

\begin{figure}
    \centering
    \includegraphics[width=0.98\columnwidth]{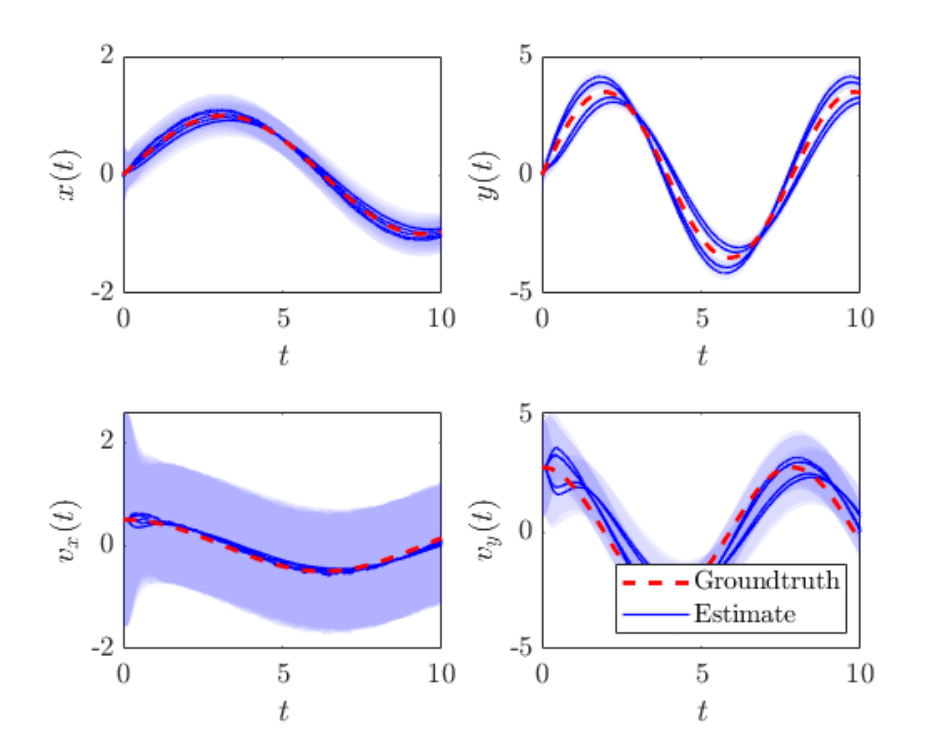}
    \caption{Estimation results from all nodes with continuous communication.}
    \label{fig:estim-fc-n5}
\end{figure}

\begin{figure}
    \centering
    \includegraphics[width=0.98\columnwidth]{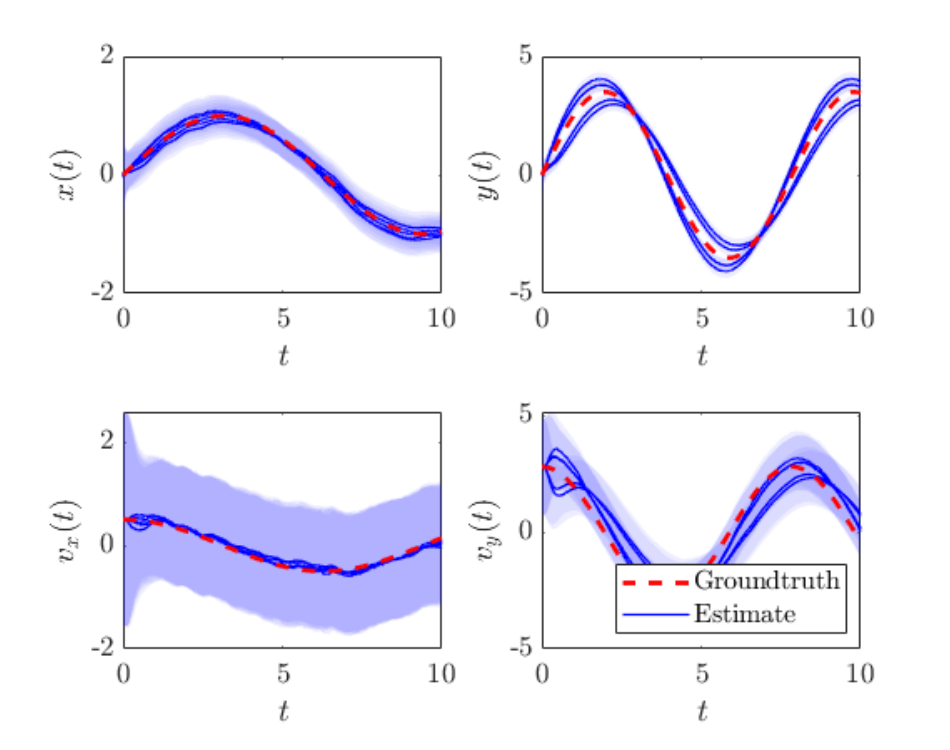}
    \caption{Estimation results from all nodes with event-triggered communication ($\delta = 25$).}
    \label{fig:estim-ev-n5-25}
\end{figure}

\begin{figure}
    \centering
    \includegraphics[width=0.98\columnwidth]{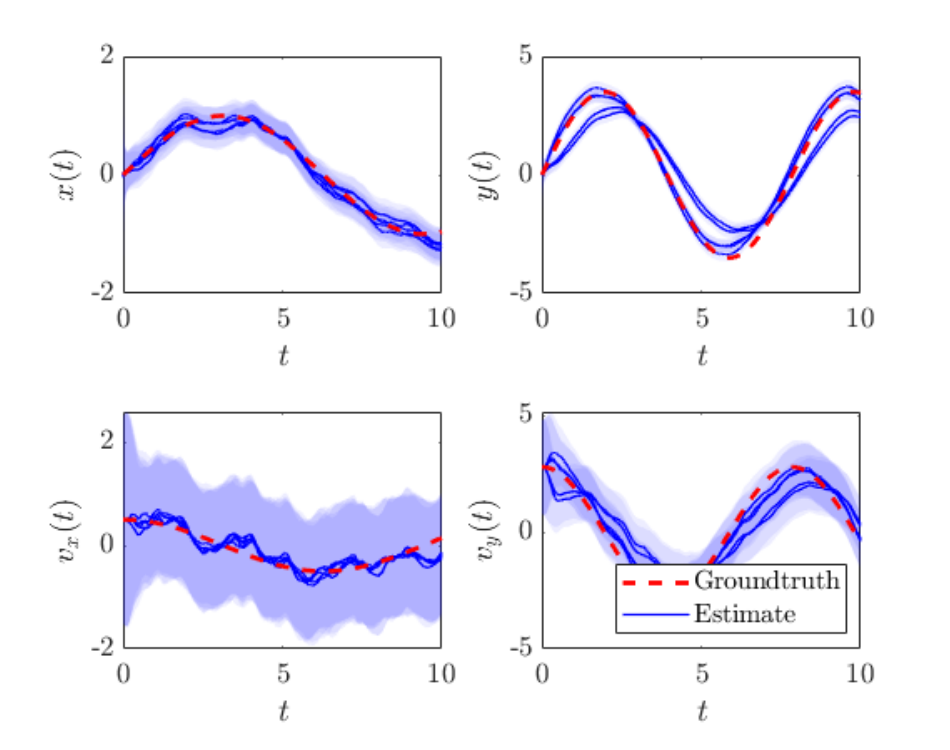}
    \caption{Estimation results from all nodes with event-triggered communication ($\delta = 50$).}
    \label{fig:estim-ev-n5-50}
\end{figure}

Figure \ref{fig:estim-fc-n5} shows the estimation results in the nodes for the ideal continuous communication case, i.e. when each node $i$ has $\hat{\mf{z}}_j(t), \, \hat{\mf{Z}}_j(t)$ available at any time. This simulation has been computed as a baseline to compare to the event-triggered case. 
Figures \ref{fig:estim-ev-n5-25} and \ref{fig:estim-ev-n5-50} show the results for the event-triggered setup for $\delta_i = 25$ and $\delta_i = 50$. It can be observed that the estimates are similar to the continuous communication case, increasing the estimation error with the triggering threshold $\delta_i$. 

\begin{figure}
    \centering
    \includegraphics[width=\columnwidth]{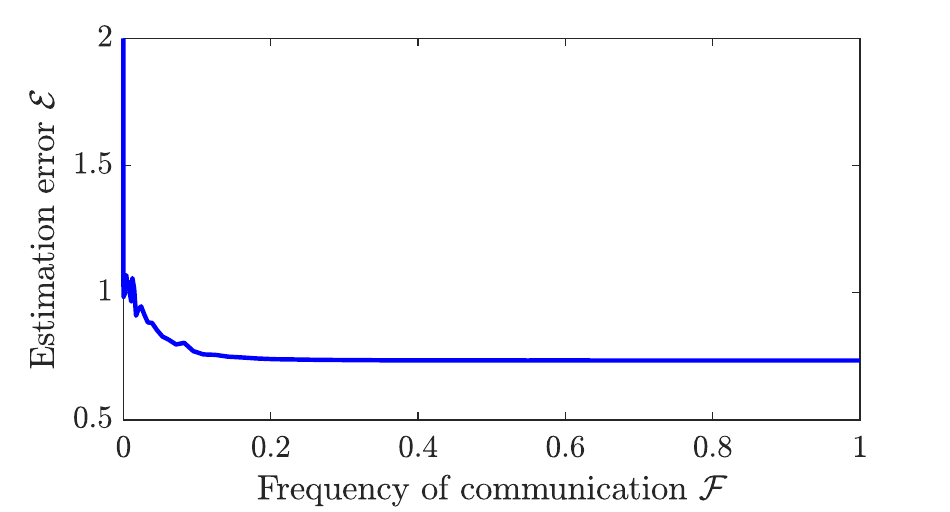}
    \caption{Trade-off between error of the estimation $\mathcal{E}$ and frequency of communication $\mathcal{F}$, normalized between 0 (no communication) and 1 (continuous communication). The frequency of communication can be reduced using our event-triggered scheme within a wide range of values, without significantly increasing the estimation error.}
    \label{fig:freq-vs-err}
\end{figure}

Lower values of $\delta_i$ provide smaller errors at the cost of an increase in frequency of communication between nodes, as is generally expected in event-triggered systems \citep{Wu2013}. This trade-off is shown in Figure \ref{fig:freq-vs-err}, which depicts the estimation error against the frequency of communication. Note that we are able to greatly reduce the frequency of communication between nodes without a significant increase in estimation error. To obtain these results, we have run simulations with the same plant as described above and different values of $\delta_i$ in a range of $[0, 80]$. Due to the stochastic nature of the problem, $S=20$ simulations have been executed for every $\delta_i$. The average estimation error and average frequency of communication of the nodes for each simulation have been computed as 
\begin{equation}
    \mathcal{E}_s = \frac{1}{N T_f} \sum_{i=1}^{N} \int_{0}^{T_f} \|\hat{\mf{x}}_i(t) - \mf{x}(t)\| \d t, \,\,\,
    \mathcal{F}_s = \frac{\sum_{i=1}^N e_i}{NT_f}
\end{equation}
where $T_f$ is the total time for the experiment and $e_i$ represents the number of events triggered in node $i$. Then, the values are averaged to obtain $\mathcal{E}$ and $\mathcal{F}$ for each $\delta_i$:
\begin{equation}
    \mathcal{E} = \frac{\sum_{s=1}^S \mathcal{E}_{s}}{S} , \,\,\, 
    \mathcal{F} = \frac{\sum_{s=1}^S \mathcal{F}_{s}}{S} 
\end{equation}
Moreover, the frequency of communication $\mathcal{F}$ is shown normalized in Figure \ref{fig:freq-vs-err}, so that 1 means continuous communication between nodes (an event is triggered at every simulation step) and 0 means no communication.


\section{Conclusions}
We have presented an approach to distributed state estimation over sensor networks for continuous-time systems, via dynamic consensus of measurements under event-triggered communication. Our method uses discrete communication between nodes, due to the triggering mechanism, but still obtains a continuous estimate. 
We have shown that applying an event-triggering mechanism to decide when each node broadcasts its local information allows to reduce communication between nodes without significantly increasing the estimation error with respect to the ideal case with continuous communication. Moreover, we have shown that the consensus error is bounded, and that the error tolerance can be tuned according to the desired performance of communication rate and estimation error.

\bibliography{bibliography.bib}

\end{document}